\documentstyle[epsfig,longtable,cite]{aipproc}

\begin{document}
\title{A model for predation pressure\\in colonial birds}

\author{J. L. Tella$^*$, M. A. R. de Cara$^{*,\dagger}$, O. Pla$^\dagger$
and F. Guinea$^{\dagger}$}
\address{$^*$ Estaci\'on Biol\'ogica de Do\~nana,
Consejo Superior de Investigaciones Cient{\'\i}ficas,
Avda. M$^{\underline{a}}$ Luisa s/n,
E-41013 Sevilla, Spain \\
$\dagger$ Instituto de Ciencia de Materiales de Madrid,
Consejo Superior de Investigaciones Cient{\'\i}ficas,
Cantoblanco, E-28049 Madrid, Spain}

\maketitle

\begin{abstract}
Different explanations have been proposed for the
existence of colonial breeding behavior in birds, but field studies
offer no conclusive results. We analyze the interplay between 
learning habits and predation pressure by means of numerical
simulations. Our analysis suggests that extremely simple learning
processes and equally simplistic models of predation pressure lead
to the formation of stable colonies. 
\end{abstract}

\section*{Introduction}
Colonial breeding behavior in birds has been extensively 
studied\cite{L68,SK90,Retal98}. Different hypotheses have been
put forward in order to explain this behavior, like minimizing the
distance required for foraging\cite{H68}, observation of conspecific 
foraging groups\cite{T56}, information transfer at the colony\cite{WZ73},
shortage of nests\cite{S76,SC87}, or predation pressure\cite{L68}.

One of the difficulties in verifying the previous hypotheses is that
present day conditions need not coincide with those which lead to
colonial behavior in the first place. Thus, modelling of bird populations
using reasonable assumptions for bird behavior can be useful in 
the elucidation of possible scenarios favorable towards
the evolution of coloniality. 

Some theoretical studies give
support to the hypothesis that information transfer at the 
colony increases the tendency towards 
colony formation\cite{BL88,BS95,B97}.

Predation can induce colonial habits in many ways. The simplest
passive mechanism is the dilution effect provided by a colony
of sufficiently large size\cite{V77,B81,T96}. In addition, the detection
and defense capabilities are enhanced 
in colonies\cite{K64,HS76,V77,A95,T96}.
On the other hand, the lack of significant predation pressure on
some colonial species has been used as evidence against the 
predation hypothesis\cite{B81,F89}, although a phylogenetic analysis
of coloniality across bird species shows a strong correlation with
exposure to predation\cite{Retal98}. 

The present work analyzes the role of predation on the formation
of colonial habits by means of a mathematical model which incoporates
some of the known facts about the response of birds to
attacks by predators, and makes simple assumptions about the memory
and learning processes at play. 
\section*{The Model.}
We assume that the available choices to birds is limited to two possibilities
each breeding
season: they can either form an individual nest, or join an existing colony. 
This binary restriction allows us to make use of the extensive literature
on competition and collective behavior of agents
with bounded rationality~\cite{A94,Z98}.

Successful breeding individuals
tend to be faithful to their previous nesting site.
Birds choose a colony or an isolated nest depending on their previous
experience.
This experience is modelled in a similar way to that used in the so called
``minority game''~\cite{Z98,Cetal99}. Each bird has, as already mentioned,
two strategies. Each of these strategies has a score, which reflects 
the degree of reproductive success that the bird would have had if it had
followed it. In this sense, the model departs from the minority game 
usually analyzed in the literature, where the score of strategies is the
same for all agents. The present version reflects the diversity of
birds, due to genetic differences and other sources, and resembles the
``individual'' version of the minority game~\cite{Cetal00}.

A bird who made a succesful choice
assigns the achieved
score to its available strategies whose outcome is the same choice
that it has taken. On the other hand, a bird whose nest is predated increases
the score of those
strategies whose outcome is the opposite.

Individual strategies, when succesful, obtain a
larger score than colonial strategies (2 vs. 1),
reflecting
the ancestral tendency of birds to solitary breeding
in the absence
of predation.
The number of sites is much larger than the number of
birds. Unless otherwise stated, the predation probability, $p_{pr}$ fluctuates
between 0 and a given value, $p_{max}$. In this way, we model
the natural variability of the abundance of predators.

At each time step, which corresponds to one breeding season, there is a finite
probability that a nest will be predated. Predation of 
nests in a colony of size $N$  is reduced by a factor proportional to $N$.
We do not follow the detailed evolution of 
individual colonies in the habitat. Instead, we assume that there is a fixed
distribution of colony sizes, such that the probability of finding a
colony of a given size is inversely proportional to its size.
This is the expected
behavior if the relative fluctuation of colony sizes is random. The distribution
is normalized to the total number of birds that have
colonial behavior.

We model the birds' finite lifespan  by setting the score of the
strategies of a given bird to zero with some finite probability,
$p=\frac{1}{v}$,
at each time step. This implies that a bird has been replaced by a new one 
with no previous experience. The total population is kept constant. 
\section*{Results.}
\begin{figure}[b!] 
\centerline{\epsfig{file=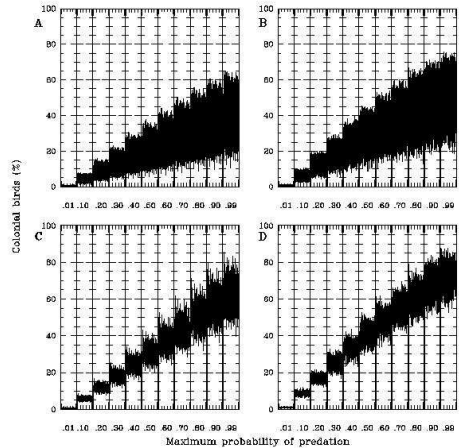,height=3.5in}}
\vspace{10pt}
\caption{Evolution of the number of colonial birds, as function of time, for
different maximum predation pressure.
Top graphs: Short-lived species relying on (A) their long-term previous
breeding experience, and on (B) their immediate previous breeding experience.
Bottom graphs: Long-lived species relying on (c) their long-term previous
breeding experience, and on (D) their immediate previous breeding experience.
%
}
\label{fig1}
\end{figure}

We show, in fig.~(\ref{fig1}), the number of colonial birds, as function of
time, for different values of $p_{max}$, and two lifespans (3 and 10 years).
The number of birds is 10000.
In the absence of predation, individual strategies acquired
two points per season,
while colonial behavior was awarded one point, reflecting the genetic
tendency towards individual nests.

Finally, we have considered two possible behaviors: a) each individual
may check the scores of all its strategies,
and uses the one with the highest score every breeding season.
Thus, birds would rely its decisions on
their long-term previous experience ({\it long-term experience}).
b) If the nest of a given individual was not predateds during
the previous season, it repeats the
last choice. 
If the nest was predated, the individual acts as in behavior (a).
 
All birds were located in individual nests at $t = 0$, and we have followed
the evolution of the population during 10$^4$ time steps. 
\section*{Conclusions.}
Our results suggest that colonial behavior can arise as a natural response
to predation pressure. Note that we made a number of conservative assumptions, in order
to avoid any bias towards colonial behavior: i) The birds have an innate tendency
towards preferring individual nests, ii) The only protection provided by the
colony is the dilution effect, iii) The distribution of colonies is such that
small colonies are more abundant, and iv) predation pressure fluctuates strongly
from year to year, allowing for the existence of periods of low predation.

The number of
colonial birds increases with increasing lifespan,
as birds accumulate experience for a longer period. This evidence is
in agreement with the observation that birds make use of their long term
breeding experiences\cite{Setal85,BG89,TH89,Oetal99}.

\end{document}